\documentstyle[aps,multicol,epsf,epsfig]{revtex}
\newcommand{\be}{\begin{equation}}
\newcommand{\ee}{\end{equation}}
\newcommand{\bea}{\begin{eqnarray}}
\newcommand{\eea}{\end{eqnarray}}
\def\sig{{\boldmath$ \sigma$}}

\begin{document}

\title{  Protein Ground  State Candidates in a Simple Model:
 An Exact Enumeration Study}
\author{ V. Shahrezaei$^{1,2}$,
N. Hamedani$^{1,2}$ and  M.R. Ejtehadi$^1$}
%\address{$^1$ {\it Institute for studies in Theoretical Physics and
%Mathematics,
% Tehran  P.O. Box 19395-5531, Iran.}\\
%$^2$ {\it Department of Physics, Sharif University of Technology,
% Tehran P.O. Box: 11365-9161, Iran.}\\} 

\maketitle \begin{abstract} The concept of the reduced set of contact maps
is introduced. Using this concept we find the ground state candidates for
Hydrophobic-Polar lattice model on a two dimensional square lattice. Using
these results we exactly enumerate the native states of all proteins for a
wide range of energy parameters. In this way, we show that there are
some
sequences, which have an absolute native state. Moreover, we study the
scale-dependence of the number of the members of the reduced set, the
number of ground state candidates, and the number of perfectly stable
sequences by comparing the results for sequences with lengths of 6 up to
20. 
\end{abstract}

\begin{multicols}{2}
%%%%%%%%%%%%%%%%%%%%%%%%%%%%%%%%%%%%%%%

\section{Introduction} The proteins are bio-macromolecules, which are made
from thousands of atoms. These atoms are in interaction with each other
and water molecules, which surround them. Basically, to determine the
states of a protein one needs to solve the problem with standard quantum
mechanical calculations, however the complexity of these macromolecules
renders this impossible. A feasible approach to this problem is based on a
coarse-grained view. In this viewpoint the proteins are made from 20 types
of monomers (amino acids). The most important point in this approach is
the determination of the effective interactions between the amino acids
\cite{MJ}.  It seems that, the information about effective inter-monomer
interaction energy and the coding of the amino acids in the sequence is
sufficient to determine the protein characteristics. 

The structural information for protein structures can be coded in a
contact map \cite{Lifson}. A contact map is a binary $L \times L$ matrix
$C$. The element $c_{ij}$ of this matrix is nonzero if $i$th and $j$th
monomers are in contact. The contact may be defined in  several ways. It
is obvious that the information coded in contact maps is not sufficient
for a complete characterization of the spatial configuration. The
short-range nature of
inter-monomer interactions suggests that, one can determine the
configuration energy in terms of contacts. Thus if one knows the
effective inter-monomer interactions in this coarse-grained approximation,
the contact
maps have sufficient information to calculate the configuration
energy. There are
many papers, which study the thermodynamical and structural
properties
of proteins, by using contact maps \cite{Vend1}.

It is well known that the biological functionality of proteins depends on
the shape of their native states. The native structure is the unique
minimum free energy structure for the protein sequence \cite{Anfinsen}.
As any protein in nature must have a well-defined function,
the uniqueness of
native states is a biological necessity for these molecules of life.
Thus searching the configuration space to find native states, by  using
the Monte Carlo methods \cite{Monte} or exact enumerations
\cite{Chan,Li,Self1} has been the
subject of many papers.
In most of previous works, the problem was studied for given
values of inter-monomer energy parameters. As our knowledge about 
the effective interactions is not certain, and the native structures of
proteins may be sensitive to these parameters \cite{Grosberg}, looking at 
the native states
for different energy parameters is relevant \cite{Self1,Shakh2}. By using
a simple
Hydrophobic-Polar (HP) lattice model we have shown in a
recent work \cite{Self2} that the number of ground state candidates for
any
sequence is unexpectedly small. This suggests that the problem can be
studied for a wide range of interaction parameters by exact
enumeration.
We study this problem on a two dimensional square lattice. In this
approach a
protein structure is modeled by a self avoiding walk on the lattice,
and, any pair of monomers which are nearest neighbors and
are not adjacent according to sequence (non-sequential neighbor) are in
contact. 

The number of possible configurations for an $L$-mer is equal to the
number of  self-avoiding walks ($N_{\rm SAW}$) with $L-1$ steps. We have :  
\be
N_{\rm SAW} \sim L^\gamma z_{\rm eff}^L,
\label{e1.0} 
\ee
in which $\gamma $ is a dimension-dependent constant, and $z_{\rm eff}$
is
the effective coordination number. For a two-dimensional square lattice,
$\gamma = {43\over 32}$ and $z_{\rm eff}= 2.64$ \cite{SAW}. Since many of
these walks give the same contact matrix, the number of possible
contact
matrices (physical maps) $N_c$, is much smaller, although it is still
very large. In a recent work \cite{Vendr} the number of physical maps was
fit to a formula similar to equation 
\ref{e1.0} and a value of $z_c= 2.29$ was obtained. 

If one is interested only in the 
native structure of proteins, the set of the contact maps can be reduced 
further, by removing all maps, which have no chance
to be a native state. We call the remaining maps, the {\it reduced
set of contact maps}.  Indeed,
this
reduction is due to the physical fact that all
effective interactions between amino acids are negative \cite{MJ}. This 
reduced set of contact maps can be used in
enumeration studies to find the possible ground states and the native
states of proteins.
In this paper we use a simple HP lattice model, to address the
problem for proteins with various lengths in  
more detail. We obtain some ground state candidates that possess some
known properties common to real proteins. Also a  stability against the
variation of interaction parameters is shown. Some evidences for
this stability has
been reported in some other works \cite{Shakh3}. 

%%%%%%%%%%%%%%%%%%%%%%%%%%%%%%%%%%%%%%%%%%%%%%%%%%%%%%%

\section{The reduced contact maps}

The effective potential energies between the 20 types of amino acids
can be described by a $20
\times 20$ interaction matrix  \cite{MJ}. The energy of a given sequence
\sig in any structure can be determined from 
\be 
E= \sum_{i,j} c_{ij} m_{\sigma_i \sigma_j}.  
\label{e2.0} 
\ee 
The $c_{ij}$ and $m_{ij}$ are respectively the elements of the contact
matrix ($C$) and
the interaction matrix ($M$). This shows that, all
configurations, which
have the same contact map, have equal energies. If we look at the energy
spectrum of one sequence, the states corresponding to such maps are
degenerate. We call such degeneracies, type-one degeneracies to distinguish
them from other kinds of 
degeneracies, which we shall encounter later \cite{Self2}. If the
energy of a sequence is minimum in such states, this
sequence does not have a unique native state. Such sequences are not
protein-like. The states corresponding to such {\it degenerate contact
maps} can never be a native state, however, we cannot exclude them from
our search, because they compete with other maps.
On the other hand, there are some maps, which cannot be the ground state,
and do not have a role in the competition for the ground state. To see
that, consider two contact matrices $C_1$ and $C_2$ 
and their subtraction ($C'=C_1 -C_2$). We call $C_2$, a {\it component} of
$C_1$
if all elements of $C'$ are non-negative ($c'_{ij}= 0$ or $1$). Note that 
$C'$ has at least one non-zero element. Using equation \ref{e2.0}, the
energy of an arbitrary sequence \sig in the configuration(s) corresponding
to the map $C_1$ can be written as:
\bea 
E_1 &=& \sum_{i,j} c_{1,ij} m_{\sigma_i \sigma_j} \nonumber \\ 
 &=& \sum_{i,j} c_{2,ij} m_{\sigma_i \sigma_j} + \sum_{i,j} c'_{ij}
m_{\sigma_i \sigma_j} \nonumber \\ 
 &=& E_2 + \sum_{i,j} c'_{ij} m_{\sigma_i \sigma_j}. \label{e5.0} 
\eea 
According to experimental data all elements of interaction matrix $M$ are
negative \cite{MJ}. Thus the second term in the rhs gives a negative
contribution to energy, and, $E_1 < E_2$, for any sequence. Then map
$C_2$ can never be a ground state. One can find all component maps such
as $C_2$, and remove them from the set of contact maps. Indeed such
component maps are related to configurations, which can fold to more
compact shapes without losing any of their old contacts. By this
procedure, a reduced collection of maps is found. We call this collection,
the {\it reduced set of contact maps}, and we represent the number of its
elements by $N_r$. We have enumerated $N_r$ for sequences with lengths up
to 20. The results are shown in figure 1. In this figure the number of
reduced maps ($N_r$) are compared with the number of self-avoiding walks
($N_{\rm saw}$) and the number of physical maps ($N_c$), on a two
dimensional square lattice. Although, all of these quantities have similar
behaviors, the growth rate of $N_r$ is very slower than the others.  If we
fit the data to equation \ref{e1.0} we obtain $\gamma_r=1.37$ and 
$z_r=2.01$. These results are not enough to see if the value of $\gamma_r$
is lattice-dependent or not. In the case of self-avoiding walks it is
lattice-independent \cite{SAW}. In figure 1 there are other points which
show the number of native states. We will discuss this matter in section
4.

Let's consider the number of contacts ($b={1\over2} \sum_{i,j} c_{i,j}$)
as a measure for the compactness of configurations. Indeed, a better
parameter is the relative compactness $\Gamma$,
\be
\Gamma= {b\over b_{\rm Max}},
\label{e6.0}
\ee
where, $b_{\rm Max}$ is the maximum number of possible contacts for
sequences of the same length. The maximum of contacts $b_{\rm Max}$, for
sequences of length 6, 8, 10, 12, 14, 16, 18 and 20 are 2, 3, 4, 6, 7, 9,
10 and 12, respectively. In figure 2, the number
of the members of the reduced set of contact maps {\it vs.} the number of
contacts is compared with corresponding number of SAWs and physical maps
for proteins of length 18. We see that the reduced set of contact maps
contains only highly compact configurations. This shows why the results of
studies on compact structure spaces are reasonable. 
In figure 3, the average compactness for SAWs, physical maps and reduced
maps are compared for sequences of various lengths.
As one can see $\langle \Gamma_{\rm SAW} \rangle <
\langle \Gamma_c \rangle < \langle \Gamma_r \rangle $. There is an
oscillatory behavior in the graphs. Note that the $b_{\rm Max}$ is an
integer. The highest ratio of $ b_{\rm Max}$ to length ($L$), is for
sequences can be fitted to a square structure. Thus, sequences with such
lengths have lower average compactness. This is due to the finite size
effect and also the fact that the number of contacts has to be an
integer; the same behavior can be observed in our other results in this
paper too.

If one scales the number of reduced maps ($N_r$) by the number of total
structures ($N_{\rm SAW}$) at each compactness, a scale-independent 
behavior can be seen (figure 4). It also seems that there is a critical
compactness, below which the compactness of the members of reduced set
never drops. We do not have an exact analytical proof, but it seems from
these data that a transition occurs in the number of reduced maps, near
the compactness of $0.8$ and it vanishes for a compactness below $0.5$.

Contact maps can correspond to more than one structure. We call such
maps, {\it degenerate maps}. These maps can not correspond the native 
state of any sequence. Within the set of reduced maps there are fewer of 
such degenerate maps, than within the set of physical maps. Figure 5
compares the percentage of non-degenerate maps for reduced and physical
maps. It seems that both of them approach asymptotic values.

%%%%%%%%%%%%%%%%%%%%%%%%%%%%%%%%%%%%%%%%%%%%%%%%%%%%%%%
\section{The ground state candidates for HP model}
The native states of proteins are to be found among the
structures corresponding to the reduced set of contact maps. The
sequence of the amino acids along the protein chain and their
interactions have an essential role in the selection of a particular
structure as the native state. In the coarse-grained viewpoint, the
the interaction between the amino acids is characterized by the 
effective energies. These effective interactions depend on the
properties of the solutions. A relevant question is how sensitive the
native structures are to changes in these interactions. We address
this question by enumerating the possible ground states of protein
sequences for a wide range of effective inter-monomer interaction energies.

Without any loss of generality, we use a hydrophobic-polar (HP) 
two-dimensional lattice model \cite{Chan2} in this paper. The general form
of the interactions between $H$ and $P$ monomers in an HP model can be written
as follows \cite{Self1,Li1}:
\bea
\label{e7.0}
E_{HH} &=& -2 -\gamma -E_c, \nonumber\\
E_{HP} &=& -1 - E_c, \nonumber \\
E_{PP} &=& - E_c,
\eea
where $E_{\sigma \sigma'}$ is the contact energy between monomers of 
types $\sigma$ and $\sigma'$. These potential energies are
only between non-sequential nearest neighbors. Here $\gamma$ and $E_c$ are
the mixing and compactness potentials respectively, two parameters which
are determined from experimental data. There are many publications
based on this model, and in most of them the values of $\gamma$
and $E_c$ are fixed \cite{Li1,Chan2}. Here, we consider them as two
free parameters and discuss our results in terms of them.

It has been argued that the following relations should hold between
inter-monomer energies:

\bea
\label{e8.0}
E_{HH} < E_{HP} < E_{PP}, \nonumber \\
E_{HH} + E_{PP} < 2E_{HP}.
\eea
These arguments are based on the compactness of
the native states \cite{Dill} and some calculations on $20 \times 20$
inter-monomer interaction matrix $M$  \cite{Li2}. These restrict $\gamma$
and $E_c$ to positive values ($\gamma,
E_c>0$).

At first sight, it might seem possible to arrive at any native state
for a given sequence by changing $\gamma$ and $E_c$. But when we consider
the geometrical properties of the ground state, we will find that these
parameters are not powerful enough to select any configuration as the
native state. In other words, the native states are stable against the
change of interaction parameters.

If we consider $H=-1$ for hydrophobic monomers and $P=0$ for polar
monomers, a given sequence can then be represented by a binary vector
(\sig) \cite{Self1}. The energy of this sequence in a configuration
characterized by a contact matrix $C$, can be written as:
\be
\label{e9.0}
E= -m -a\gamma- b E_c,
\ee
where $m$, $a$ and $b$ are three integers, related to \sig and $C$ as
follows:
\bea
\label{e10.0}
m&=& - {\mbox {\sig}}^{t} \cdot C \cdot{\bf 1}, \nonumber \\
a&=&  {1\over2} {\mbox{\sig}}^{t} \cdot C
 \cdot{\mbox{\sig}}, \nonumber \\
b&=& {1\over2} {\bf 1}^{t} \cdot C \cdot{\bf 1}.
\eea
It can be seen that $m$ is equal to the number of all
non-sequential neighbors of
$H$ monomers in the configuration, $a$ is the number of $H-H$ contacts and
$b$ is the number of all contacts. It can be shown that the
following inequalities hold between these parameters \cite{Nima}.
\be 
\label{e11.0}
m-b\le a\le {m\over 2} \le b.
\ee

Equation \ref{e9.0} suggests that the energy levels of a given sequence
can be described by three integer numbers  $(m,a,b)$. It is highly
probable that these states are degenerate. There are three types of
degeneracy:
\begin{itemize}
\item Type 1: $C=C'$ \\ In which case two or more configurations with
different shapes  have the same contact matrix. These configurations will
remain degenerate for any sequence, and any choice of $\gamma$ and $E_c$.
These are the configurations corresponding to the degenerate maps already  
mentioned in section 2. This type of degeneracy, is more
probable for configurations with low compactness (see figure 2). Note that
we are not talking about the
configurations which are related to each other by spatial symmetries, i.e.
rotation, reflection, etc., for our purpose such configurations are
identical.
\item Type 2: $(m,a,b)=(m',a',b')$ but $C \ne C'$ \\
In this case one particular sequence has the same $m$, $a$ and $b$ values
in two or more configurations. This degeneracy persists for any value of
$\gamma$ and $E_c$, but may disappear for another sequence. Although, this
degeneracy depends on sequence coding, the  $b=b'$ condition is
purely geometrical, and is a necessary condition for this degeneracy. 
\item Type 3: $E=E'$, but $(m,a,b) \ne (m',a',b')$ \\ 
One sequence has the same energy in two different states $(m,a,b)$ and
$(m',a',b')$, provided that $\gamma$ and $E_c$ obey the following
relation: 
\be
\label{e12.0}
(m-m') + (a-a')\gamma + (b-b') E_c = 0.
\ee
This degeneracy is related to both sequence coding \sig and inter-monomer
interactions.
\end{itemize}

The first type of these degeneracies is completely geometric. The second
one depends on both geometry and the amino acids' coding sequence. These
two types do not depend on the values of the interaction energies. Thus,
in the energy spectrum of any sequence there are some states, which are
degenerate independently from the potential. If the ground state of a
particular sequence is one of these degenerate states, that sequence does
not have a unique native structure.

The third type is not actually a degeneracy at all.  Equation \ref{e12.0}
corresponds to a line in the parameter space of $E_c$ and $\gamma$. This
line is a level crossing line. Degeneracy actually occurs only on the
line, and a highly accurate fine-tuning is needed to reach a point on this
line. For the two sets of interaction energy parameters on the two sides
of this line, the energy ordering of the states is different. For any pair
of states such an ordering line exists. By drawing all ordering lines in
the space of $E_c$ and $\gamma$, this space is divided into many ordering
zones. We are only interested in the ground state, which means that many
of these ordering lines are not relevant. Some of them only govern the
ordering of the excited states. By removing the irrelevant lines, one gets
a diagram which shows the ground state cells (Fig. 6). As mentioned before
changing the inter-monomer interaction parameters inside any of these
cells does not change the ground state. In some recent works \cite{Mourik}
this picture is introduced to show the stability of native states against
change in the interaction parameters \cite{note}. They only looked at one
of these cells in the neighborhood of some selected interaction values. 
But by looking at the whole energy space, one can find all possible ground
states and their corresponding cells. Any such cell in the space of energy
parameters is associated with one ground state candidate. The number of
cells is equal to the number of ground state candidates ($G_c($\sig$)$). 
By drawing such diagrams, one can easily find the ground state for any
choice of $E_c$ and $\gamma$. Fig. 6 shows this diagram for a 20-mer. In
this example there are only seven possible ground states. The cells marked
with the numbers ``1" and ``2" correspond to type-1 and type-2 degenerate
states respectively, therefore there is no unique native structure for
these cells. The sequence in this example has 3 non-degenerate states.
These structures are shown in the figure. It is possible that all the
ground state candidates of a given sequence are degenerate. These
sequences constitute universally bad sequences i.e. for any set of
interaction parameter values they do not have a native structure. Any
sequence which is not a bad sequence, we call a {\it good sequence}.
Nearly $54\%$ of the sequences of length 20 are good sequences, i.e. for
some specific set of energy parameters they have a native state.

The interesting point in figure 6 is that the number of ground state
candidates is very small. The largest value of $G_c$, for sequences with
length $6, 8, 10, 12, 14, 16, 18, 20$ are $1, 1, 1, 3, 4, 5, 6, 7$
respectively. Fig. 7 shows the histogram of $G_c($\sig$)$ for all
sequences with $L=20$. The light gray area in this figure shows the result
for all $2^{20}$ sequences, and the dark area shows the results for good
ones. From this diagram it can be seen that the mean value of
$G_c($\sig$)$ is very small. The average of $G_c($\sig$)$ for various
lengths is shown in figure 8. However, the data in hand is not enough to
draw a reliable conclusion about the number of ground state candidates for
sequences of large length, but the average number does not seem to grow
very rapidly, and the growth rate appears to be linear.  Extrapolating the
growth rate to sequences of length 200, 30 ground state candidates is
predicted on average. Comparison of the average value of $G_c($\sig$)$ for
these sequences with the number of all configurations (i.e. for sequences
with length 20 the number of sequences is on the order of $10^8$), shows
that the geometric constraints play an important role in selecting a state
as the ground state. The reason that there are few ground state candidates
for any sequence can be given by a geometrical argument \cite{Self2}. 
This argument shows that the upper estimate for maximum $G_c$ is $L^2$. 

As figure 8 shows, there are some good sequences with $G_c=1$. This means
that for any set of energy parameter values, they have the same unique
ground
state. Fig. 9 shows some of these sequences and their unique native
structures. Indeed the native states of these sequences have perfect
stability with respect to a change of the energy parameters. Our
enumeration shows that these {\it absolute native structures} are to be
found among the most compact structures. As figure 10 shows, although the
ratio of the number of {\it perfectly stable sequences} to the number of
all possible proteins decreases with increasing $L$, their actual number
increases. This suggests that for the proteins with typical lengths near
that of natural proteins, perfectly stable sequences constitute a small
but non-zero fraction of all possible sequences. A relevant question is
whether the existence of these perfectly stable sequences is due to the
simplifications in our model.  Actually we can not give an exact answer to
this question, but such sequences may exist in models with more monomer
types.

The existence of these sequences may answer some questions about protein
folding. Their number is small compared with the huge number of the
possible amino-acids sequences, their native states are highly compact and
are stable against the changes in the inter-monomer interactions
(i.e the properties of the solution). 

%%%%%%%%%%%%%%%%%%%%%%%%%%%%%%%%%%%%%%%%%%%%%%%%%

\section{Native structures}

In section 2 we introduced the reduced set of contact maps. As it was
shown the number of maps belonging to this set $N_r$, is very smaller than
number of structures $N_{\rm SAW}$. But the number of those structures
which can be the native state, is still much less. The number of possible
native structures, $N_{\rm native}$, is shown in figure 1. In this figure
all those structures which have been the native state of some sequence for
at least one set of energy parameter values, have been counted.  Fitting
the data on a equation similar to equation \ref{e1.0}, gives $\gamma_{\rm
native}=1.87$ and $z_{\rm native}=1.68$. In figure 2, we have compared the
number of native structures as a function of their compactness 
with the total number of physical maps and with the number of maps in 
the reduced set for $L=18$. It can be seen that there are no native
structures with less than 8 contacts. Also the average compactness of
native states is compared in figure 3.

We can introduce a designability parameter $D$ for these native states. 
However our definition is a bit different from the commonly used
definition \cite{Li}.  According to the common definition, designability
shows how many times a structure is selected as the native state for a
fixed set of interaction parameters. In our case we count how many times a
structure becomes the candidate for a non-degenerate ground state.

Figure 11 shows the histogram of designability for structures with length
20. As one can see the results are very similar to those for a fixed set
of energy parameters in the space of compact structures \cite{Self1,Li}.
The average designability as a function of compactness for $L=20$ is shown
in figure 12. As the diagram shows the peak average designability occurs 
for the most compact structures and it falls sharply with decreasing
compactness. Thus if one is only interested in highly designable
structures, it is reasonable to search the space of compact structures.

%%%%%%%%%%%%%%%%%%%%%%%%%%%%%%%%%%%%

\section{The space of Energy parameters, $E_c$ and $\gamma$}

One of the important aspects of the work done in this paper, is that we
can find the exact results for any range of energy parameters. The time it
takes for this program to find the ground state candidates for all
sequences by exact enumeration, is on the same order as that of the usual
enumeration schemes for only one particular set of energy parameters.
Because the average number of ground state candidates is very small, the
determination of the native ground states for any range of interest only
takes a little time. We found the native states of all sequences of length
$20$, for all pairs of energy parameters within a $12 \times 12$ square in
arbitrary units, with a grid size of $0.1$ (14400 points). The number of
protein-like sequences (sequences which have unique ground states) is
shown in figure 13. As one can see, there are jumps in the number of
protein-like sequences. These jumps specify the borders of regions of
relative stability within the space of energy parameters.  A closer
examination shows that these border lines contain sharp dips adjacent to
the jumps. The large changes in the number of protein-like sequences shows
that when we cross these borders the ground states of many sequences
change, and the degenerate ground states are replaced by non-degenerate
ones (or vice versa). However, nothing can be said about the details of
these changes. One can get some idea about what is happening on these
border lines by comparing the contour plot for figure 13.a (figure 13.b)
with the ordering lines diagram for one particular sequence (figure 6). As
mentioned in section 3, the ordering lines specify level crossings and
type-3 degeneracies only occur on the ordering line itself. These ordering
lines constitute the underlying cause of the sharp dips observed in the
borders. This is more evident in figure 14. In this figure we have shown
those points in the energy parameter space where at least one type-3
degeneracy occurs. This diagram is in fact a superposition of diagrams
like figure 6, for all sequences, and any line in it corresponds to many
ordering lines between ground state candidates cells. 

We can find similar information for other types of degeneracies. For
example, the number of sequences which have type-1 degenerate ground
states, is shown in figure 15. As one can see in this diagram, the number
of such sequences vanish for large $E_c$ and small $\gamma$. For large
$E_c$ the number of contacts $b$ plays an essential role in the selection
of the ground state (equation \ref{e9.0}). Type-1 degeneracies do not
occur for highly compact sequences (see figure 2). Thus this type of
degeneracy is more relevant in the region $E_c < \gamma$. We have not
shown the corresponding information for type-2 degeneracies as they
contain no new information, similar border jumps can be observed in the
number of sequences with this type of degeneracy too. The maximum
percentage of sequences with non-degenerate, type-1 degenerate, and type-2
degenerate ground states in the chosen region are $40.0\%$, $5.06\%$ and
$64.9\%$,
respectively. 

In addition to obtaining information about the sequences, with this
procedure also finds the ground states. Since the energy parameters
determine which states are the ground states, the number of structures
which can be the native state of some particular sequence also depends on
the energy parameters. Figure 16 shows the number of native states as a
function of the energy parameters. The importance of compactness at for
large values of $E_c$ can also be seen in this diagram.  Note that the
smallest value for the number of native states is 503.  This number
corresponds to the number of most compact structures of length 20. Again,
large jumps in the number of native states are observed. One can also find
the average designability of the structures by dividing the data of
figures 13 and 16 (the ratio of the number of sequences to corresponding
number of native structures).

\section{Conclusion}

Due to the short-range nature of inter-monomer interactions, the
configuration energy of protein sequences can be determined by using
configuration contact matrices. In this paper, it has been shown that for
this class of problems, where one is interested in native states of
proteins, the space of physical contact maps can be reduced to a very
smaller set by removing all irrelevant maps. We have found the reduced set
of contact maps for sequences of lengths up to $20$ in this paper by exact
enumeration. This reduced set of contact maps shows a scale-independent
behavior as shown in figure 4.

Using the reduced set of contact maps, the ground state candidates for all
sequences were found in the HP model. The number of these ground state
candidates is quite small. The ground state candidates divide the space of
energy parameters into several cells. By finding this cell structure for
all sequences, we have found the native states for all sequences of
different lengths, for a wide range of energy parameters. Jumps are
observed in the number of protein-like sequences. These jumps are related
to boundaries of the aforementioned cells. 

Another interesting result is that we find some sequences with absolute
native states i.e. their native states are not sensitive to the values of
energy parameters. Our results show that the number of such perfectly
stable sequences grows with length, however, their percentage decreases. 

Because the key tool used in this paper has been the structural information
contained in the contact maps, the qualitative results can be generalized
to all contact models, regardless of the details of the lattice and the
contact rules. 

{\bf Acknowledgement:} We would like to thank  S.E. Faez,
R. Gerami,  R. Golestanian, A.Yu. Grosberg,  N. Heydari, M. Khorami, S.
Rouhani and H. Seyed-Allaei for helpful comments.

\end{multicols}
%%%%%%%%%%%%%%%%%%%%%%%%%%%%%%%%%%%%%%
\newcommand{\PNAS}[1]{ Pros.\ Natl.\ Acad.\ Sci.\ USA\ {\bf #1}}
\newcommand{\JCP}[1]{ J.\ Chem.\ Phys.\ {\bf #1}}
\newcommand{\PRL}[1]{ Phys.\ Rev.\ Lett.\ {\bf #1}}
\newcommand{\PRE}[1]{ Phys.\ Rev.\ E\ {\bf #1}}
\newcommand{\JPA}[1]{ J.\ Phys.\ A\ {\bf #1}}

%\end{multicols}
%%%%%%%%%%%%%%%%%%%%%%%%%%%%%%%%%%%%%%%%%%%%%%%%%%%%%%%%%%%%%%%%%%%%%%%%%%%
\newpage
\begin{center}
\Large
Figure Captions \\[15mm]
\normalsize
\end{center}

{\bf Figure 1.}

\parbox{16cm}{The number of self avoiding walk structures,
	physical contact maps, reduced set of contact maps and native
	structures, 
	{\it vs.} length of sequences.
	}

{\bf Figure 2.}

\parbox{16cm}{
	Distribution of the number of structures {\it vs.} the number of
	contacts for sequences of length $18$.
	}

{\bf Figure 3.}

\parbox{16cm}{
	The average compactness of structures for SAW, physical maps,
	reduced maps and native structures.
	}

{\bf Figure 4.}

\parbox{16cm}{
	The number of reduced maps that scaled by the number of all
	structures at each compactness, for sequences with length 8 to 20.
	There is a transition near to $0.8$ and a cut off near to $0.5$.
	The later can be seen
	better in logarithmic scale (inner graph).
	}

{\bf Figure 5.}

\parbox{16cm}{
	The percentage of non-degenerate maps for reduced and physical
	maps.
	}

{\bf Figure 6.}

\parbox{16cm}{
	The space of energy parameters for sequence $HPPPHPHPHPPHPHPHPHHP$ 
	is divided to six cells.
	The integer numbers $(m,a,b)$, inside any cell indicate the
	ground state 
	corresponding to the cells. Three of  these states are degenerate.
	The types of degeneracies for degenerate states and shape of structures
	for non-degenerates are indicated in the cells.}

{\bf Figure 7.}

\parbox{16cm}{ 
	The histogram of the number of ground state candidates for 20-mers.
	The light and dark gray area show the results for all and good
	sequences respectively. There are some ``good sequences" with only
	one ground state candidate.}

{\bf Figure 8.}

\parbox{16cm}{ 
	The average of the number of ground state candidates for all and
	good sequences {\it vs.} length of sequences.}

{\bf Figure 9.}

\parbox{16cm}{
	Four example for perfectly stable sequences and their absolute
	native structures.
	For any positive value of $\gamma$ and $E_c$ these sequences are
	folded
	uniquely in the shown structures. }

{\bf Figure 10.}

\parbox{16cm}{
	The ratio of the numbers of perfectly stable sequences to all
	sequences
	decreases with length of sequences, but their absolute numbers
	increase (inner graph). }

{\bf Figure 11.}

\parbox{16cm}{
	The histogram of number of structures with a given designability.
	}

{\bf Figure 12.}

\parbox{16cm}{
	The average designability for structures with a given number of
	contacts, for $L=20$.}

{\bf Figure 13.}

\parbox{16cm}{
	The number of protein-like sequences of length 20, for given
values of energy 
parameters in a $12 \times 12$ square region (arbitrary units); a) Three
dimensional plot, b) Contour plot. }

{\bf Figure 14.}

\parbox{16cm}{
	The points in the energy parameter space (arbitrary units),
	where type-3 degeneracies occur, for sequences of length 20. The
grid size is $0.1$.}

{\bf Figure 15.}

\parbox{16cm}{
	The number of sequences  of
length 20, with type-1 degenerate ground states, for given
values of energy 
parameters in a $12 \times 12$ square region (arbitrary units).}

{\bf Figure 16.}

\parbox{16cm}{
	The number of native states for sequences  of
length 20, for given
values of energy 
parameters in a $12 \times 12$ square region (arbitrary units).}

\vspace{5mm}


\begin{thebibliography}{40}
\bibitem{MJ}{S. Miyazawa and A. Jernigan, Macromolecules {\bf 18},
	534 (1985).}
\bibitem{Lifson}{S. Lifson and C. Sander, Nature {\bf 282}, 109 (1979).}
\bibitem{Vend1}{For a recent review, see M. Vendruscolo and E. Domany,
cond-mat/9901215.}
\bibitem{Anfinsen}{C.B. Anfinsen, Science {\bf 181}, 223 (1973).}
\bibitem{Monte}{C. Camacho and D. Thirumalai, \PNAS{90}, 6369 (1993); A.M. 
	Gutin, V.I. Abkevich and E.I. Shakhnovich, \PRL{77}, 5433 (1996).}
\bibitem{Chan}{H.S. Chan and K.A. Dill, \JCP{95}, 3775 (1990).} 
\bibitem{Li}{H. Li, R. Helling, C. Tang and N. Wingreen ,
	     Science {\bf 273}, 666 (1996).}
\bibitem{Self1}{M.R. Ejtehadi, N. Hamedani, H. Seyed-Allaei,
	       V. Shahrezaei and M. Yahyanejad,
	       \PRE{57}, 3298 (1998); M.R. Ejtehadi, N. Hamedani, H.
	Seyed-Allaei, V. Shahrezaei and M. Yahyanejad,
	       \JPA{31}, 6141 (1998).}
\bibitem{Grosberg}{V.S. Pande, A.Yu. Grosberg and T. Tanaka,
	     \JCP{103}, 9482 (1995).}
\bibitem{Shakh2}{E.L. Kussell and E.I. Shakhnovich, Preprint
	Cond-mat/9904377.}
\bibitem{Self2}{M.R. Ejtehadi, N. Hamedani and V. Shahrezaei, To appear in
	Phys. Rev. Lett.; Preprint Cond-mat/9811127.}
\bibitem{SAW}{N. Madres and G. Slade, {\it The Self-avoiding Walk},
	(Birkhauser, Boston, 1993).}
\bibitem{Vendr}{M. Vendruscolo, B. Subramanian, I. Kanter,
	        E. Domany and  J. Lebowitz, Preprint Cond-mat/9810285.}
\bibitem{Shakh3}{V.I. Abkevich, A.M. Gutin and E.I. Shakhnovich,
Biochemistry {\bf 33}, 10026 (1994); E.I. Shakhnovich, V.I. Abkevich and
O.B. Ptitsyn, Nature {\bf 379}, 96 (1996).}
\bibitem{Chan2}{H.S. Chan,
K.A. Dill, \JCP{90}, 492 (1989);
	    H.S. Chan, K.A. Dill, D. Shottle,
	    ``Statistical Mechanics and Protein Folding",
	    {\it  Prinston Lectures on Biophysics}, W. Bialek ed.,
	    (World Scientific, 1992).}
\bibitem{Li1}{R. Melin, H. Li, N. S. Wingreen and C. Tang,
	     Preprint Cond-mat/9806197.}
\bibitem{Dill}{K.A. Dill, Biochemistry {\bf 24}, 1510 (1985);
	    K.A. Dill, S. Bromberg, K. Yue, K.M. Fiebig,
	    D.P. Yee, P.D. Thomas and H.S. Chan,
	    Protein Science {\bf 4}, 561 (1995).}
\bibitem{Li2}{H. Li, C. Tang and N. Wingreen ,
	     Phys. Rev. Lett. {\bf 79}, 765 (1997).}
\bibitem{Nima}{ N. Hamedani, V. Shahrezaei and M.R. Ejtehadi, In
Preparation.}
\bibitem{Mourik}{J. Mourik, C. Clementi, A. Maritan, F. Seno and J. R.
	Banavar,
	    Preprint Cond-mat/9801137; C. Clementi, A. Maritan 
	    and J. R. Banavar, \PRL{81}, 3287 (1998); F. Seno, A. Maritan
	    and J. R. Banavar, Proteins {\bf 30}, 244 (1998). }
\bibitem{note}{In their work they parameterize the energy space
	    by $E_{HP}$ and $E_{PP}$ instead of  $\gamma$ and $E_c$.}

\end{thebibliography}
 \end{document}